\documentclass[fleqn,usenatbib]{mnras}

\usepackage{newtxtext,newtxmath}

\usepackage[T1]{fontenc}

\DeclareRobustCommand{\VAN}[3]{#2}
\let\VANthebibliography\thebibliography
\def\thebibliography{\DeclareRobustCommand{\VAN}[3]{##3}\VANthebibliography}

\usepackage{graphicx}	
\usepackage{amsmath}	
\usepackage{orcidlink}
\usepackage{threeparttable}
\usepackage{graphicx}
\usepackage{subcaption}

\usepackage{xcolor}

\title{Suppression of star formation at the centre of barred AGN galaxies}

\author[T. Yamamoto et al.]{Takashi Yamamoto \orcidlink{0000-0003-3124-0210},$^{1}$\thanks{E-mail: yb.office@icloud.com}
Nario Kuno \orcidlink{0000-0002-1234-8229},$^{2,3}$
Daisuke Iono \orcidlink{0000-0002-2364-0823},$^{4}$
Takuya Hashimoto \orcidlink{0000-0002-0898-4038},$^{2,3}$
Sophia K. Stuber \orcidlink{0000-0002-9333-387X},$^{4,5}$
\newauthor Daizhong Liu \orcidlink{0000-0001-9773-7479},$^{6}$
Thomas G. Williams \orcidlink{0000-0002-0012-2142},$^{7}$
and Akio K. Inoue \orcidlink{0000-0002-7779-8677}$^{1,8}$
\\
$^{1}$Waseda Research Institute for Science and Engineering, Faculty of Science and Engineering, Waseda University, 3-4-1 Okubo, Shinjuku, Tokyo 169-8555, Japan\\
$^{2}$University of Tsukuba, 1-1-1 Tennoudai, Tsukuba, Ibaraki 305-8577, Japan\\
$^{3}$Tomonaga Center for the History of the Universe (TCHoU), Faculty of Pure and Applied Sciences, University of Tsukuba, Tsukuba, Ibaraki 305-8571, Japan\\
$^{4}$National Astronomical Observatory of Japan, 2-21-1 Osawa, Mitaka, Tokyo 181-8588, Japan\\
$^{5}$Max Planck Institute for Radio Astronomy, Auf dem Hügel 69, 53121 Bonn, Germany\\
$^{6}$Purple Mountain Observatory, Chinese Academy of Sciences, 10 Yuanhua Road, Nanjing 210023, China\\
$^{7}$ UK ALMA Regional Centre Node, Jodrell Bank Centre for Astrophysics, Department of Physics and Astronomy, The University of Manchester, Oxford Road,\\ Manchester M13 9PL, UK\\
$^{8}$Department of Physics, School of Advanced Science and Engineering, Faculty of Science and Engineering, Waseda University, 3-4-1 Okubo, Shinjuku,\\ Tokyo 169-8555, Japan\\
}

\date{Accepted XXX. Received YYY; in original form ZZZ}

\pubyear{2026}

\begin{document}
\label{firstpage}
\pagerange{\pageref{firstpage}--\pageref{lastpage}}
\maketitle

\begin{abstract}

We measured the Concentration ($C$) index of H$\ \alpha$ and CO~(\textit{J}\ =\ 2–1) in 17 nearby star-forming galaxies from the PHANGS survey. We have found four barred spiral galaxies with a $C$(H$\ \alpha$)/$C$(CO) ratio of $\sim$ 0.3–0.4, while the other 13 galaxies exhibit ratios of $\sim$ 1 (ranging from 0.7 to 1.3). All four barred galaxies with low ratios host active galactic nuclei (AGNs), consistent with a scenario in which star formation is suppressed by mechanical and radiative feedback from the AGN. Therefore, negative feedback may be effective in these four barred galaxies even when AGN activity is relatively weak. The formation of bar structures causes molecular gas to collect in the central region, leading to starburst activity. However, after the starburst, the remaining gas becomes inefficient for star formation rapidly due to AGN feedback. This can mean the quenching process occurs more rapidly in AGN-barred galaxies. Furthermore, since gas remains in the central region, AGN activity is likely to continue. These quenching processes are a unique mechanism found in barred spiral galaxies and are essential to understanding galaxy evolution.

\end{abstract}




\section{Introduction}

Despite the vigorous study of nearby galaxies in recent years, the evolutionary process of galaxies is still unclear. The most common scenario is that multiple mergers of disc galaxies form massive elliptical galaxies~\citep[e.g.,][]{2006MNRAS.373.1013C, 2008ApJS..175..356H}. 
The merger of gas-rich galaxies triggers starbursts~\citep[e.g.,][]{jog1992triggering,gao2001molecular}. However, we must not overlook that the formation of galactic bar structures efficiently transports molecular gas toward the centre, thereby inducing central starbursts. For example, according to a recent investigation, \citet{2022MNRAS.513.2850B} utilized N-body/hydrodynamical simulations to illustrate the gas concentration at the centre due to the galactic bar formation, and the subsequent central starburst and formation of the galactic bulge. On the other hand,~\citet{maeda2023statistical} reported that the star formation in the bar region except for galactic centre is systematically suppressed, caused by dynamical effects, such as strong shocks, shears, and fast cloud-cloud collisions. 

Furthermore, recent studies have revealed a growing number of barred spiral galaxies at $z\sim2$ through observations with the James Webb Space Telescope~\citep[\textit{JWST}, e.g.,][]{guo2023first, huang2025large}, and at $z=3.1$~\citep{umehata2025adf22}, suggesting that such dynamically mature structures were already present when the Universe was only a few billion years old. The role of barred structures in galaxy evolution extends back to the most star-forming period in the cosmic star formation history.

The purpose of this study is to clarify the role of galactic bars in galaxy evolution using the Concentration index ($C$). We focus on bar structures because they not only trigger starbursts but may also lead to the quenching of star formation. 
We also investigate the role of active galactic nuclei (AGNs), given that molecular gas preferentially accumulates in the central regions of barred spiral galaxies \citep[e.g.,][]{sakamoto1999bar,sheth2005secular,kuno2007nobeyama,yamamoto2025quantitative}, and that AGN feedback can suppress star formation in galactic centres \citep[e.g.,][]{lammers2023active}.
\citet{lammers2023active} compared the star formation trends of 279 low-redshift AGN galaxies with 558 inactive control galaxies from the Sloan Digital Sky Survey-IV (SDSS-IV) Mapping Nearby Galaxies (MaNGA) at Apache Point Observatory survey. From their statistical investigation, they have shown that the AGN galaxies have significantly suppressed central (kpc-scale) star formation rates (SFRs), lying up to a factor of two below those of the control galaxies, providing direct observational evidence of AGN feedback suppressing star formation. 

This paper is organized as follows: Section 2 describes the data for the 19 galaxies used in this paper and explains the CAS parameters. Section 3 presents the key results revealed by this analysis, and discusses the results in detail, and Section 4 is a summary.

\section{Data and analysis method}

\subsection{PHANGS-ALMA archival data}
We retrieved The Physics at High Angular Resolution in Nearby GalaxieS (PHANGS)-ALMA (the Atacama Large Millimeter/submillimeter Array) archive data v4.0\footnote{https://sites.google.com/view/phangs/home/data} from the Canadian Advanced Network for Astronomical Research\footnote{https://canfar.net/} (CANFAR) consortium in Canadian Astronomy Data Centre (CADC). The survey data were taken as a Large Program in ALMA Cycle 5 (2017.1.00886L, PI: E.Schinnerer). The Public Release contains CO~(\textit{J}\ =\ 2--1), hereafter CO~(2--1), data for 74 star-forming, massive and close to face-on galaxies ALMA-visible within 20~Mpc, including galaxies characterized by grand design, flocculent, lenticular, and strongly barred morphologies~\citep[e.g.,][]{stuber2023gas}. The processing of the PHANGS-ALMA survey and its pipelines are discussed in ~\citet{2021ApJS..255...19L}. The main sample images were constructed with data from the 12-m array, the 7-m array, and the Total Power Array, meaning that all spatial frequencies were observed and recovered. PHANGS-ALMA delivers `cloud scale' observations, a high-resolution view of the molecular gas clouds with a resolution of $\sim100$ pc.  Massive giant molecular clouds are often found to have radii of $\sim$ 30 – 60 pc~\citep[e.g.,][]{2021ApJS..257...43L}. That is, the resolution achieved by PHANGS–ALMA is sufficient to resolve the typical size of individual giant molecular clouds.

In PHANGS-ALMA, there are two types of data products: the strict mask products and the broad mask products. The broad mask products have more completeness and cover a larger fraction of the total flux than the strict mask products~\citep{2021ApJS..255...19L}.
\citet{2021ApJS..255...19L} recommend using the strict mask for calculations sensitive to noise, e.g., many types of kinematic analysis, and the broad mask for any analysis aimed at a complete characterization of the emission. Completeness is crucial to this paper’s analysis, which measures the concentration of CO and H$\ \alpha$. Therefore, we use the broad mask products for the analysis in this study.

\subsection{PHANGS-MUSE archival data}
The PHANGS-MUSE survey is a large program that uses the VLT/MUSE integral field spectrograph to map 19 massive ($9.4<\log(M_*/M_{\odot})<11.0$) nearby ($D \lesssim20\ \mathrm{Mpc}$) star-forming disc galaxies (Program IDs: 1100.B-0651/PI: E. Schinnerer; 095.C-0473/PI: G. Blanc; and 094.C-0623/PI: K. Kreckel)~\citep{2022A&A...659A.191E}. In this paper, we analyse 19 PHANGS-MUSE galaxies. PHANGS-MUSE delivers a `cloud scale' high-resolution view of the ionised interstellar medium with a median resolution of 50 pc. PHANGS-MUSE survey archival data was released as DR1.0 in 2021. We retrieved DR1.0 from CANFAR consortium in CADC. We utilized H$\ \alpha$ and H$\ \beta$ emission line data products (MAPS) from DR1.0. 

The data from PHANGS-ALMA and PHANGS-MUSE can be compared because they cover almost the same regions of the same galaxies \citep[see Fig. 4 of][]{2022A&A...659A.191E}. To ensure consistency and enable a fair comparison across the entire dataset, we convolved the data resolution to a physical scale of $\sim 180$ pc, which is the lowest resolution galaxy (NGC 1672) in PHANGS-ALMA.

\subsection{Features of the 19 PHANGS-MUSE galaxies.}

The distribution of optical morphological types of the sample based on the Third Reference Catalogue of Bright Galaxies~\citep[RC3;][]{1991rc3..book.....D} is as follows: three SA galaxies, nine SB galaxies, and seven SAB galaxies. S0 and peculiar galaxies are not included in this sample. Of these galaxies, NGC 4254, NGC 4303, NGC 4321, and NGC 4535 are member galaxies of the Virgo cluster. The galaxies considered isolated are NGC 628, NGC 1087, and NGC 5068. IC 5332 is considered a galaxy toward a group of galaxies. The remaining galaxies belong to some cluster or group of galaxies. The PHANGS-ALMA sample includes a variety of morphological types, such as barred spiral galaxies, non-barred spiral galaxies, lenticular galaxies, and peculiar galaxies. In contrast, the PHANGS-MUSE sample comprised only barred and non-barred spiral galaxies. Moreover, within the PHANGS-ALMA and PHANGS-MUSE samples, the proportion of non-barred spiral galaxies among all spiral galaxies is approximately 25 per cent and 15  per cent, respectively. However, since this paper focuses on barred spiral galaxies, this is not a major issue.

For quantitative morphological analysis, we must account for the presence of an AGN at the centre. At least eight of these 19 galaxies host AGNs, so the central regions affected by AGN emission must be masked when conducting the quantitative morphological analysis. NGC 1365, NGC 1433, NGC 1566, NGC 1672, NGC 3627, NGC 4303, and NGC 7496 are classified as Seyfert galaxies, while NGC 4535 is classified as a LINER. Among the Seyfert galaxies, only NGC 1566 is a Seyfert 1; the others are Seyfert 2~\citep[e.g., NASA/IPAC Extragalactic Database\footnote{https://ned.ipac.caltech.edu};][]{combes2013alma,belfiore2023calibrating}. In addition, \cite{2005AJ....130...73H} reported that they have found evidence of a weak AGN in NGC 1300. Even in the Baldwin–Phillips–Terlevich (BPT) diagnosis ~\citep[e.g.,][]{2003MNRAS.346.1055K, 2006MNRAS.372..961K, 2019A&A...622A.146M} results within 1 kpc from galactic centre, NGC 1300 is positioned near the low-ionisation nuclear emission-line region galaxies (LINERs) with a $\log_{10}$(\ion{O}{iii} $\lambda5007$ / H$\ \beta$) of $-0.17$, $\log_{10}$(\ion{N}{ii} $\lambda6583$ / H$\ \alpha$) of $-0.22$. NGC 4321 has been reported as a transient object: \ion{H}{ii}/LINER \citep{2005A&A...441.1011G}. However, this galaxy does not lie in the AGN area on the BPT diagnosis, according to  $\log_{10}$(\ion{O}{iii} $\lambda5007$ / H$\ \beta$) of $-0.64$, $\log_{10}$(\ion{N}{ii} $\lambda6583$ / H$\ \alpha$) of $-0.46$. It is important to note that this sample has a high proportion of AGN galaxies ($\sim$ 50 per cent). For all of these AGN galaxies, including NGC 1300, we make the masking within the range of the point spread function (PSF) at the centre of the galaxy due to the extremely high H$\ \alpha$ brightness. Accordingly, we make the same masking of the CO~(2--1) emission in these galaxies. Additionally, to evaluate whether this masking operation affects the analysis results, we also perform the analysis without masking.

\subsection{Modified CAS indices system}

We applied CAS parameters~\citep{2003ApJS..147....1C}, which are often used to analyse stellar light distribution, to molecular gas and \ion{H}{ii} regions. For CAS index measurements, all galaxies were convolved to a resolution of 180 pc to account for variations in spatial resolution among the sample. We employed the same method used in~\citet{yamamoto2025quantitative}. The CAS parameters were calculated for each galaxy using a Python script written specifically for this purpose and calling libraries to aid the analysis:  
\textsc{Casa}\footnote{https://casa.nrao.edu}~\citep{bean2022casa}, \textsc{Carta}\footnote{https://cartavis.org/}~\citep{comrie2021carta},
\textsc{NumPy}\footnote{https://numpy.org}~\citep{harris2020array}, \textsc{SciPy}\footnote{https://scipy.org}~\citep{2020NatMe..17..261V}, \textsc{Astropy}\footnote{https://www.astropy.org}~\citep{2018AJ....156..123A}, and \textsc{Matplotlib}\footnote{https://matplotlib.org}~\citep{2021zndo...5706396C}. The library for advanced image processing \textsc{OpenCV}\footnote{https://opencv.org}~\citep{bradski2008learning} was used for calculating the Asymmetry index.

We define the Concentration ($C$) as the logarithm of the ratio of the radius containing 90 per cent of the total flux ($r_{90\%}$) to the radius containing 10 per cent of the total flux ($r_{10\%}$). Namely Concentration ($C$) is calculated using the equation~(\ref{eq1}) below.
\begin{equation}
    C = 5 \times \log(r_{90\%}/r_{10\%}),
\label{eq1}
\end{equation}
This equation is slightly different from the definition adopted by \citet{2003ApJS..147....1C}, which uses the radius containing 80\% of the total flux ($r_{80\%}$) to 20\% of the total flux ($r_{20\%}$) instead of $r_{90\%}$ to $r_{10\%}$.
In this study, we utilize PHANGS-ALMA and PHANGS-MUSE high-resolution data capturing fine detail; therefore, we can measure $r_{10\%}$ more correctly. The Concentration can be measured more accurately by using dynamic ranges of $r_{90\%}$ and $r_{10\%}$.

The Asymmetry ($A$) is calculated by rotating the galaxy image 180 degrees around its center and then subtracting it from the original image. This is expressed as equation~(\ref{eq2}):

\begin{equation}
        A = \frac{\Sigma_{x,y=1,1}^{M,N}|I_{x,y}-R_{x,y}|}{2\ \Sigma_{x,y=1,1}^{M,N} I_{x,y}},
\label{eq2}
\end{equation}
where $I_{x,y}$ is the original image, and $R_{x,y}$ is the rotated image. $M$ and $N$ are the number of pixels of the galaxy image.

Clumpiness ($S$) is originally an index of the patchiness of stellar light distribution in a galaxy \citep{2003ApJS..147....1C}. $S$ is an index for capturing the local features of the distribution. The definition is the equation~(\ref{eq3}) below;

\begin{equation}
    S=10\times \Sigma_{x,y=1,1}^{M,N} \frac{I_{x,y}-I_{x,y}^\sigma}{I_{x,y}},
\label{eq3}
    \end{equation}

where $I_{x,y}$ is the flux value of the galaxy at position $(x,y)$, $I_{x,y}^\sigma$ is the same value obtained after reducing the resolution by a smoothing filter with width $\sigma$, and $M$ and $N$ are the number of pixels of the galaxy image. Here, we use $\sigma=500 ~\mathrm{pc}$ in this work same to ~\citet{yamamoto2025quantitative}. For details of the CAS parameters, see~\citet{yamamoto2025quantitative}.

\section{Results and discussion}

The characteristics of the 19 PHANGS-MUSE galaxies and the measured Concentration ($C$) indices are presented in Table~\ref{tab:001}.

\begin{table*}
    \centering
    \caption{Features of PHANGS-MUSE galaxies and Concentration ($C$) measurement.}
    \begin{minipage}{0.85\textwidth}
    \begin{tabular}{lccccccccc}
\hline
galaxy id & Galaxy Type & $T$ stage & AGN & Type & $R_\mathrm{bar}/R_{25}$ & bar & $C$(CO) & $C$(H$\ \alpha$) & $C$(H$\ \alpha$) / $C$(CO)\\
&&&&&&&&&(Ro$C$)\\
&(1)&(2)&(3)&(4)&(5)&(6)&(7)&(8)&(9)\\
\hline

IC 5332 & SA & $6.8\pm0.4$ & 0 & Non-barred  & -- & non &3.4&2.9& 0.9\\
NGC 628 & SA & $5.2\pm0.5$ & 0 & Non-barred & -- & non &3.1&2.4& 0.8\\
NGC 1087 & SAB & $5.2\pm0.8$ & 0 & Barred(B) & 0.19 & SF bar &5.7&4.6& 0.8\\
NGC 1300 & SB & $4.0\pm0.2$ & 1 & Barred(A) & 0.46 & prominent &7.2&2.2& 0.3\\
NGC 1365 & SB & $3.2\pm0.7$ & 1 & Barred(B) & 0.25 & bar &5.5&7.3& 1.3\\
NGC 1385 & SB & $5.9\pm0.5$ & 0 & Barred(N) & 0.08 & short &3.8&4.2& 1.1\\
NGC 1433 & SB & $1.5\pm0.7$ & 1 & Barred(B)/GV & 0.35 & prominent &6.9&6.4& 0.9\\
NGC 1512 & SB & $1.2\pm0.5$ & 0 & Barred(B)/GV & 0.28 & bar &5.6&5.7& 1.0\\
NGC 1566 & SAB & $4.0\pm0.2$ & 1 & Barred(A) & 0.16 & bar &5.4&2.3& 0.4\\
NGC 1672 & SB & $3.3\pm0.6$ & 1 & $*$ & 0.40 & prominent &6.7&$*$& $*$\\
NGC 2835 & SB & $5.0\pm0.4$ & 0 & Barred(N) & 0.11 & short &3.2&2.2& 0.7\\
NGC 3351 & SB & $3.1\pm0.4$ & 0 & Barred(B) & 0.24 & bar &6.6&6.6& 1.0\\
NGC 3627 & SB & $3.1\pm0.5$ & 1 & Barred(A) &0.21 & bar &5.0&2.0& 0.4\\
NGC 4254 & SA & $5.2\pm0.7$ & 0 & Non-barred & -- & non &4.1&3.8& 0.9\\
NGC 4303 & SAB & $4.0\pm0.1$ & 1 & Barred(A) & 0.18 & bar &5.2&2.1& 0.4\\
NGC 4321 & SAB & $4.0\pm0.3$ & 0 & Barred(B) & 0.23 & bar &5.9&6.0& 1.0\\
NGC 4535 & SAB & $5.0\pm0.4$ & 1 & $**$ & 0.19 & bar &7.0&$**$& $**$\\
NGC 5068 & SAB & $6.0\pm0.4$ & 0 & Barred(N) & 0.11 & short &3.2&3.3& 1.0\\
NGC 7496 & SB & $3.2\pm0.6$ & 1 & Barred(B) & 0.35 & prominent &7.7&6.5& 0.8\\ \hline
\end{tabular}
\vspace{2mm} 
    \label{tab:001}

{\small Column (1): Galaxy Type used the following three papers as references. \citet{1991rc3..book.....D}, \citet{2015ApJS..217...32B} and \citet{stuber2023gas}. Column (2): Hubble classification obtained from Hyper-Linked Extragalactic Databases and Archives (HyperLEDA)\footnote{http://atlas.obs-hp.fr/hyperleda/}. Column (3) shows whether or not the galaxy hosts AGN. `1' means that the galaxy hosts an AGN, and `0' means it does not (Based on \citet[][]{belfiore2023calibrating, combes2013alma, 2005AJ....130...73H} and NASA/IPAC Extragalactic Database). Regarding AGN types, see Table~\ref{tab:002}.} Column (4) displays whether the star formation in these barred spiral galaxies is Bulge(B)-type or Arm(A)-type. Barred(N) is a short-barred galaxy type similar to a non-barred galaxy's star formation distribution. Column (5): $R_\mathrm{bar}/R_{25}$ was calculated according to~\citet{yamamoto2025quantitative}, where $R_\mathrm{bar}$ is the bar length and $R_{25}$ is the isophotal radius. `Short' signifies galaxies with a short bar (Bar length ($R_\mathrm{bar}/R_{25}) \lessapprox 0.1$). `Prominent' signifies galaxies with a prominent bar ($R_\mathrm{bar}/R_{25}  \gtrapprox 0.3$).  `SF bar' is a rare bar structure with star formation. Columns (7) and (8) show Concentration values of CO~(\textit{J}\ =\ 2--1) and $\mathrm{H}\alpha$. Column (9) indicates $C$(H$\ \alpha$) to $C$(CO) ratio. $*$ mark indicates dust extinction in clump structures. $**$ mark indicates dust extinction in galactic centre. We exclude these two galaxies from the measurement of $C$(H$\ \alpha$) and Ro$C$ in our work.
\end{minipage}
\end{table*}

\begin{table*}
\centering
\caption{Calculation results for various measures of the 11 barred spiral galaxies.}
\begin{minipage}{0.9\textwidth}
\begin{tabular}{cccccccccccc}
\hline
galaxy id & $C$(CO) & $C$(H$\ \alpha$) & Ro$C$ & $ \log_{10}{M_{*}}$ & $\log_{10}\mathrm{SFR}$ & $C/T$ & $\Delta$MS & $f_\mathrm{gas}({M_\mathrm{H_2}/M_{*}})$ & Type & AGN type & Environment\\ & & & & (${\mathrm{M}_\odot}$) & $(\mathrm{M}_{\odot}\mathrm{yr^{-1}})$ & & &(\%) & & & \\
 &(1)&(2)&(3)&(4)&(5)&(6)&(7)&(8)&(9)&(10)&(11)\\
\hline
NGC~1087 & 5.7 & 4.6 & 0.8 & 9.94 & 0.11 & 0.20 & 0.33 & 10.4 & B & no & Isolated\\
NGC~1300 & 7.2 & 2.2 & 0.3 & 10.62 & 0.07 & 0.08 & -0.18 & 3.3 & A & LINER & Group \\ 
NGC~1365 & 5.5 & 7.3 & 1.3 & 11.00 & 1.24 & 0.45 & 0.72 & 13.5 & B & Sy2 & Cluster\\
NGC~1433 & 6.9 & 6.4 & 0.9 & 10.87 & 0.05 & 0.14 & -0.36 & 1.7 & B(GV) & Sy2 & Group\\
NGC~1512 & 5.6 & 5.7 & 1.0 & 10.72 & 0.11 & 0.31 & -0.21 & 1.5 & B(GV) & no & Group \\
NGC~1566 & 5.4 & 2.3 & 0.4 & 10.79 & 0.66 & 0.06 & 0.29 & 5.5 & A & Sy1 & Group\\
NGC~3351 & 6.6 & 6.6 & 1.0 & 10.37 & 0.12 & 0.53 & 0.06 & 2.5 & B & no & Group\\
NGC~3627 & 5.0 & 2.0 & 0.4 & 10.84 & 0.59 & 0.01 & 0.19 & 6.0 & A & Sy2 & Group\\
NGC~4303 & 5.2 & 2.1 & 0.4 & 10.51 & 0.73 & 0.04 & 0.54 & 13.5 & A & Sy2 & Cluster\\
NGC~4321 & 5.9 & 6.0 & 1.0 & 10.75 & 0.55 & 0.20 & 0.21 & 10.4 & B & no & Cluster\\
NGC~7496 & 7.7 & 6.5 & 0.8 & 10.00 & 0.35 & 0.24 & 0.53 & 9.3 & B & Sy2 & Group\\ \hline
\end{tabular}
\vspace{2mm}
    \label{tab:002} 
    
{\small This table lists the 11 A-type and B-type barred spiral galaxies selected from the 19 galaxies in Table 1. It excludes the two galaxies with significant dust attenuation, the three N-type galaxies with short bars, and the non-barred galaxies. Short-barred galaxies exhibit molecular gas distributions very similar to those of non-barred galaxies, so they are excluded from the discussion of typical barred galaxies (A-type, B-type).} Columns (1) and (2) show Concentration of CO~(2--1) and Concentration ($C$) of H$\ \alpha$, respectively. (3) is the ratio of $C$ (H$\ \alpha$) to $C$ (CO). (4) and (5) were taken from ~\citet{2021ApJS..257...43L}. (6) is centre-to-total flux ratio of H$\ \alpha$. Here, the centre refers to the region within a radius of 1 kpc from the galactic centre. (7) are calculated values of the SFR offset from the star-forming main sequence. Column (8) was taken from Table 3 of~\citet{yamamoto2025quantitative}. (9) displays whether the star formation in these barred spiral galaxies is bulge-type (B) or arm-type (A). (10) shows AGN types, and `no' means it does not have an AGN. See section 2.3. (11) shows each galaxy's environment; 'Isolated' means an isolated galaxy, 'Cluster' is a cluster member galaxy, and 'Group' is a group member galaxy.
\end{minipage}
\end{table*}

\subsection{Suppression of star formation at the centre}
\begin{figure*}
  \centering
\begin{subfigure}[b]{0.3\textwidth}
    \centering
    \includegraphics[width=\textwidth]{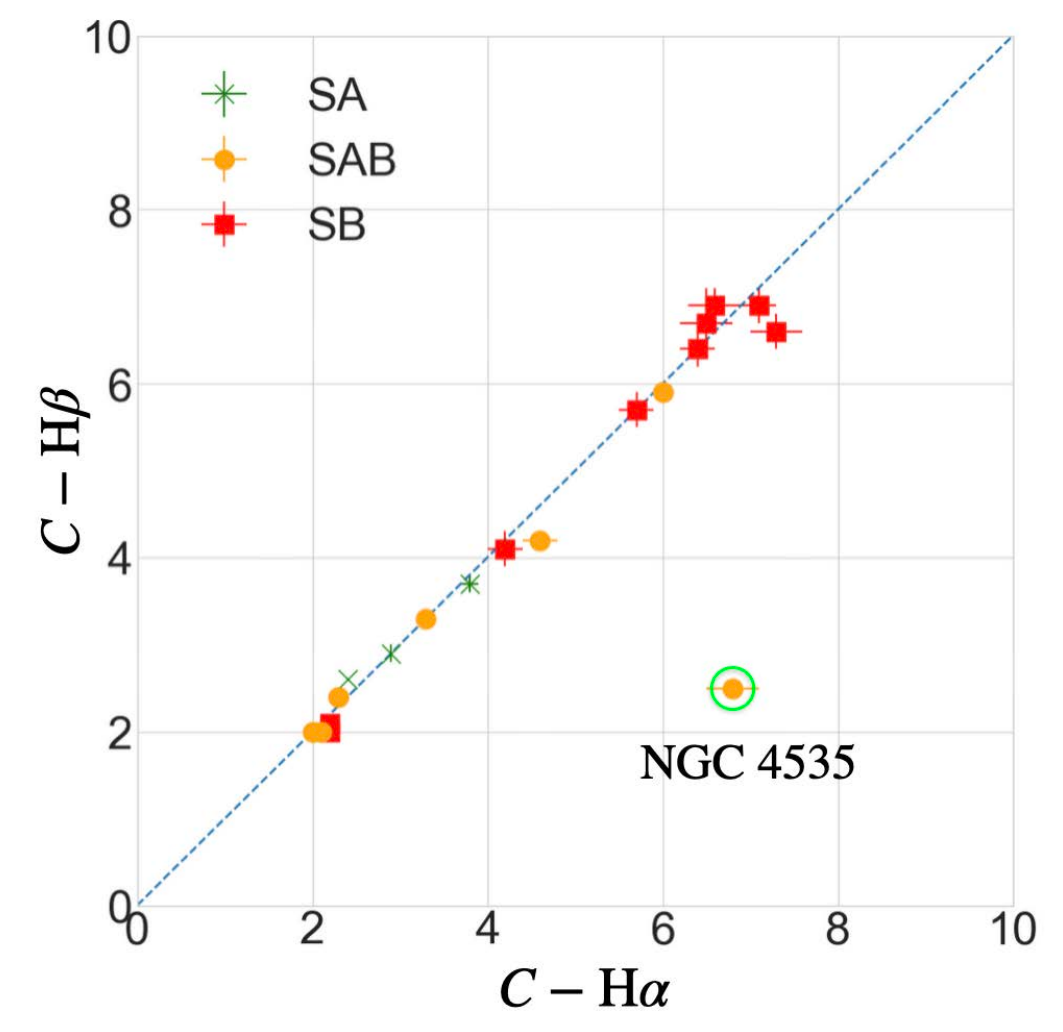}
    \caption{Concentration}
    \label{fig:panelA}
  \end{subfigure}
  \hfill 
  \begin{subfigure}[b]{0.3\textwidth}
    \centering
    \includegraphics[width=\textwidth]{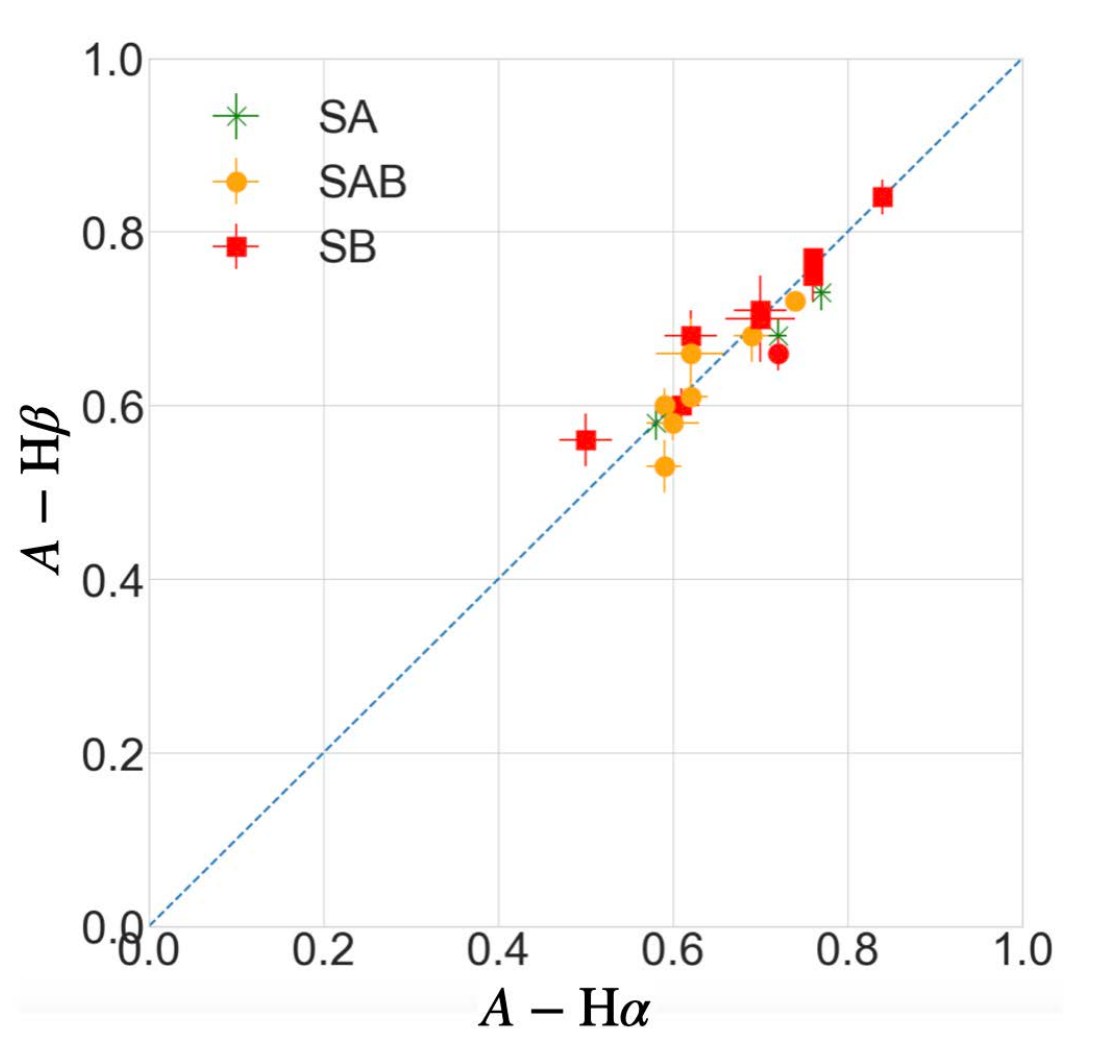}
    \caption{Asymmetry}
    \label{fig:panelB}
  \end{subfigure}
  \hfill 
  \begin{subfigure}[b]{0.3\textwidth}
    \centering
    \includegraphics[width=\textwidth]{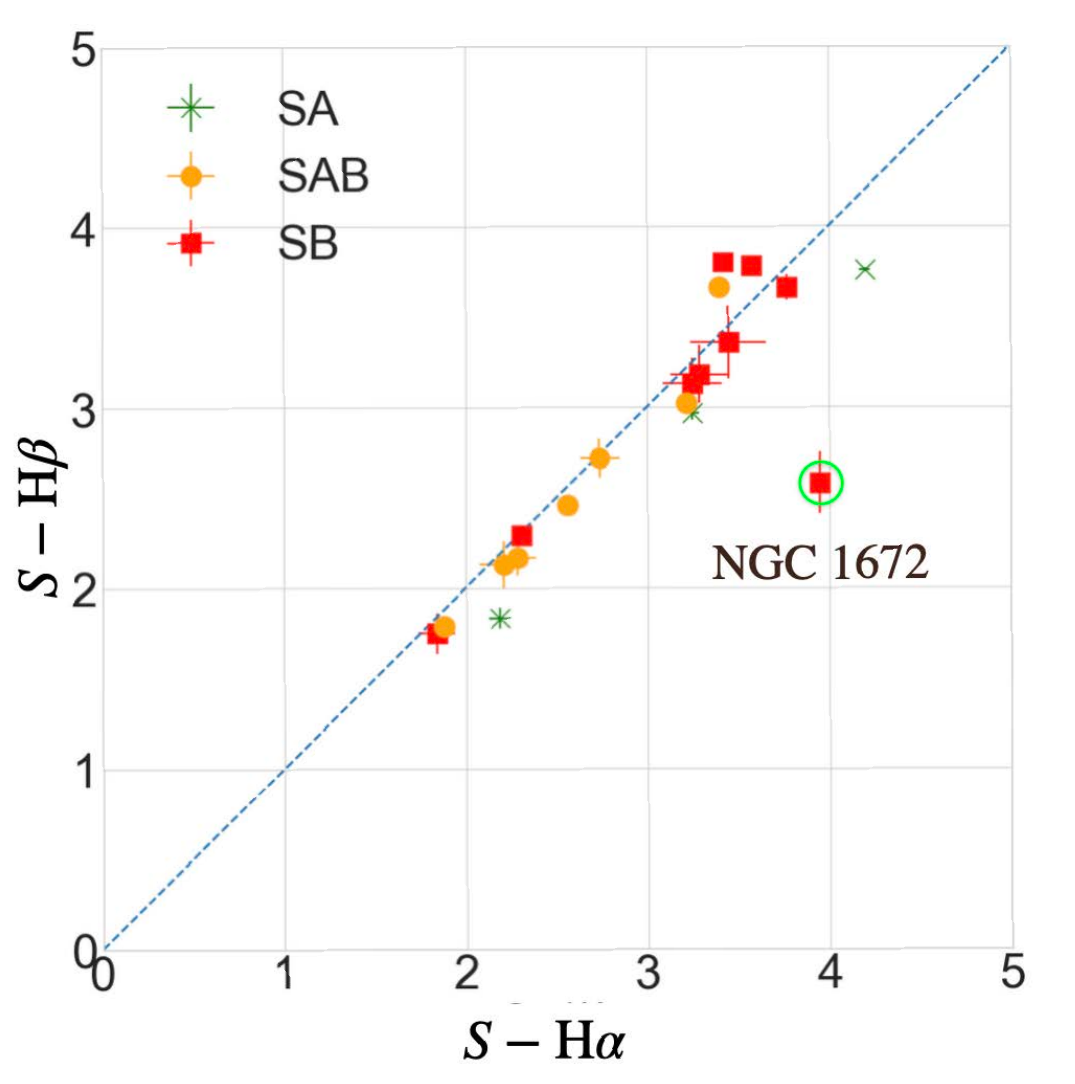}
    \caption{Clumpiness}
    \label{fig:panelC}
  \end{subfigure}
\caption{Effect of dust attenuation on the CAS parameters measured from H$\ \alpha$ and H$\ \beta$ in 19 galaxies. Panels (a), (b), and (c) show Concentration ($C$), Asymmetry ($A$), and Clumpiness ($S$), respectively. NGC 4535 in panel (a) and NGC 1672 in panel (c) show significant deviations from the one-to-one relations, while the $A$ values in panel (b) lie close to the one-to-one line. SA, SAB, and SB galaxies are represented by green crosses, orange circles, and red squares, respectively. The dotted lines indicate the one-to-one relations.}

{Alt text: The effect of dust attenuation in 19 galaxies. Three panels show the correlation of H$\ \alpha$ and H$\ \beta$ for each CAS parameter.}

\label{fig:001}

\end{figure*}

Dust attenuation can be corrected using the Balmer decrement under the case B recombination approximation, in which Lyman-series photons are assumed to be absorbed and reprocessed within the nebula, yielding a theoretically predicted intrinsic H$\ \alpha$/H$\ \beta$ ratio of approximately 2.86 for typical H II region conditions; $T_e\sim10^4$ K and low electron density~\citep{osterbrock2006astrophysics}. However, since not all nebulae necessarily follow case B, unexpected systematic errors may arise in the CAS parameters. Here, we use a more straightforward method and remove galaxies affected considerably by the attenuation. The most effective method is to create a scatter plot of the correlation between H$\ \alpha$ and H$\ \beta$ for each index of CAS. Figure~\ref{fig:001} shows the correlation between H$\ \alpha$ and H$\ \beta$ for $C$, $A$ and $S$, respectively. In this figure, the dotted lines indicate the one-to-one relation of each quantity. 
The correlation between H$\ \alpha$ and H$\ \beta$ for $A$ index values closely follows the one-to-one line. However, for the $C$ index, the value for NGC 4535 falls well below this line, because $C$(H$\ \beta$) is much lower than $C$(H$\ \alpha$). This means dust attenuation is strong in the galaxy centre. In terms of the $S$ index as well, NGC 1672 is strongly affected by dust attenuation in the clumpy structure region, because this galaxy shows that $S$(H$\ \beta$) is much lower than $S$(H$\ \alpha$). So, we excluded NGC 1672 and NGC 4535 from further study. Thus, our analysis focuses on the remaining 17 galaxies.

\begin{figure}
\begin{center}
  \includegraphics[width=\columnwidth]{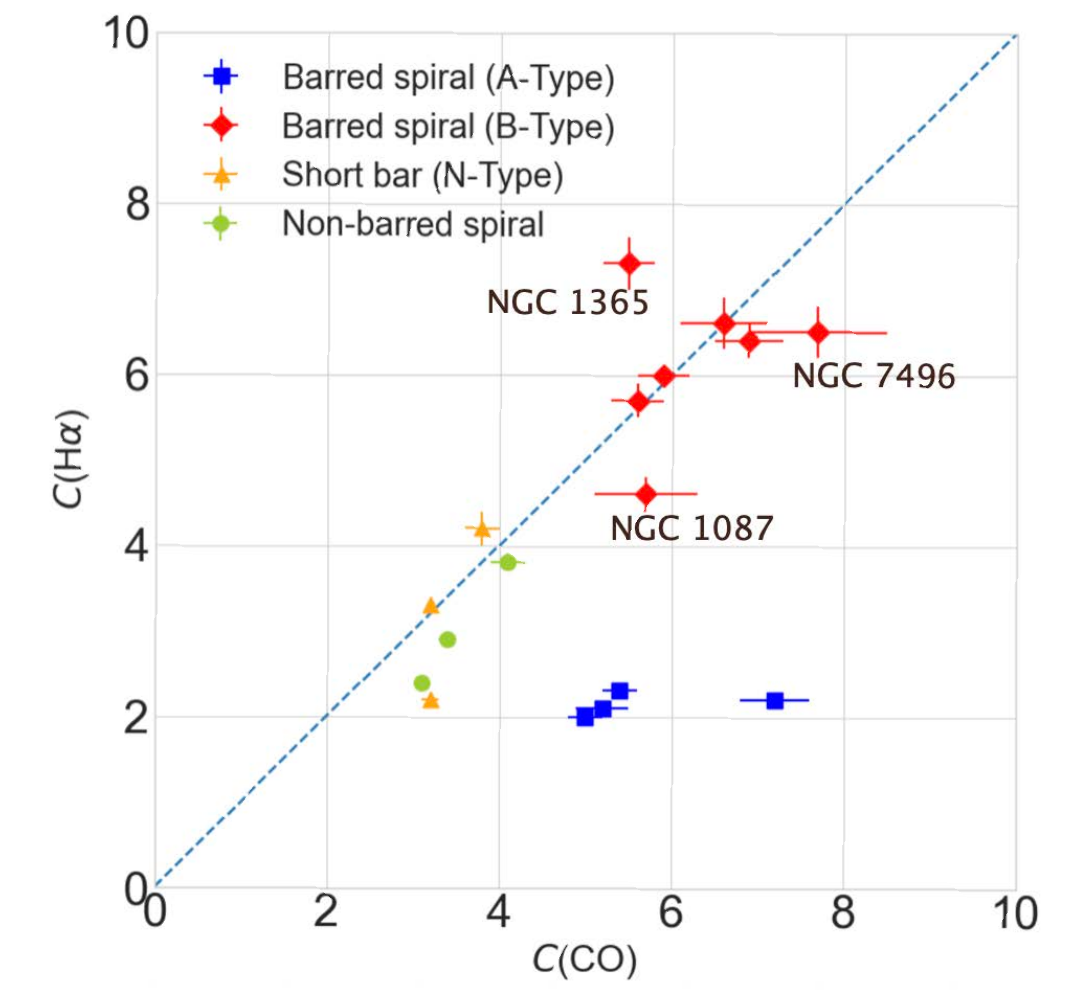}
 \end{center}
\caption{Scatter plot of $C$ (CO) versus $C$ (H$\ \alpha$) for four galaxy types: A-type barred galaxies (blue squares), B-type barred galaxies (red diamonds), N-type (short-bar) galaxies (orange triangles), and non-barred galaxies (yellow-green circles). The diagonal line indicates the one-to-one relation. All four A-type galaxies lie below this line.}

{Alt text: Scatter plot for $C$ (CO) and $C$ (H$\ \alpha$) in four types of galaxies, which are A-type barred galaxies, B-type barred galaxies, N-type (short bar) barred galaxies, and non-barred galaxies.}
 \label{fig:C_COHA2}
 
\end{figure}

\begin{figure*}
\begin{center}
  \includegraphics[width=145mm]{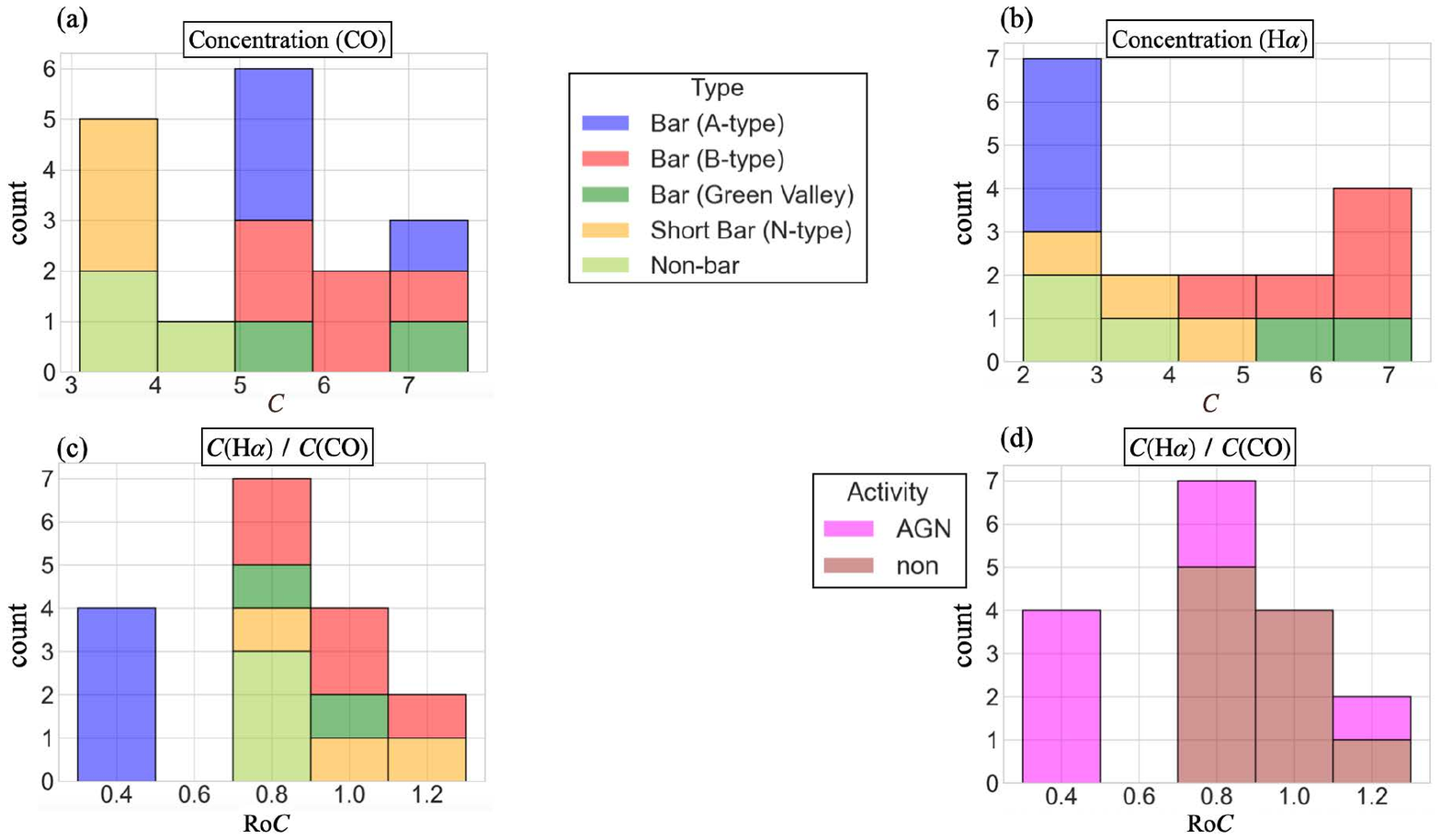}
 \end{center}
\caption{Four panels (a–d) show histograms of the Concentration ($C$) measured from CO and H$\ \alpha$ emission and their ratio (Ro$C$) for 17 galaxies. Panels (a) and (b) show C(CO) and C(H$\ \alpha$), respectively, while panels (c) and (d) show the Ro$C$ distributions by galaxy type and AGN activity, respectively. Colours indicate galaxy types in panels (a)–(c) and AGN activity in panel (d), as shown in the legends.}

{Alt text: Four panels represent histograms of $C$ of H$\ \alpha$ and CO and their ratio (Ro$C$) for each type of galaxy. Panels (a)–(c) show the distributions for different galaxy types, while panel (d) compares the Ro$C$ distributions of AGN-hosting and non-AGN galaxies.}
 \label{fig:histogram}
\end{figure*}

We measured the Concentration ($C$) index of H$\ \alpha$ and CO~(2–-1) in the 17 PHANGS-MUSE galaxies, after removing two galaxies with significant dust attenuation in optical emission line. We then calculated $C$(H$\ \alpha$)/$C$(CO) (hereafter, Ro$C$). The results are shown in Table~\ref{tab:002} and Figure~\ref{fig:C_COHA2}. Figure~\ref{fig:C_COHA2} shows that the derived $C(\mathrm{H}\alpha)$ and $C$(CO) are generally aligned near the one-to-one line, except for four barred galaxies which are located significantly below the one-to-one line (blue squares). The mean RoC values for all 17 galaxies and for the 13 galaxies excluding the four A-type galaxies are 0.8 and 0.9, respectively. Furthermore we calculated the centre-to-total flux ratio ($C/T$) of H$\ \alpha$. Here, the centre refers to the region within a radius of 1 kpc from the galactic centre. The results are shown in Table~\ref{tab:002}. The mean $C/T$ for H$\ \alpha$ of the four barred spiral galaxies is $0.05\pm0.01$. This value is six times lower than the mean of the remaining barred spiral galaxies ($0.30\pm0.05$). This indicates suppressed star formation in the central region compared to the spiral arms. For convenience, we refer the former four barred galaxies as `Arm-type' barred galaxies (A-type) and the seven barred galaxies with high Concentration ($C$) values in both~CO~(2--1) and H$\ \alpha$ as `Bulge-type' barred galaxies (B-type). 

Figure~\ref{fig:histogram} (a-d) show histograms of $C$ for CO and H$\ \alpha$, and their ratio (Ro$C$). It is seen from Figure~\ref{fig:histogram}~(a), that the distribution of $C$(CO) between barred and non-barred galaxies is different, with a mean of $5.7\pm0.3$ and $3.5\pm0.3$, respectively. We note that the $C$(CO) in all three short-barred (N-type) galaxies (bar length $R_\mathrm{bar}/R_{25} \lessapprox 0.1$, where $R_{25}$ is the isophotal radius) is low ($3.4\pm0.2$), and they are consistent with the values found in non-barred spiral galaxies (see Table~\ref{tab:001}). 
$R_\mathrm{bar}/R_{25}$ was calculated according to~\citet{yamamoto2025quantitative}. Short-barred galaxies exhibit molecular gas distributions very similar to those of non-barred galaxies, so they are excluded from the discussion of typical barred galaxies (A-type, B-type) in Section 3.2. On the other hand, it is clear from the histogram of $C$ (H$\ \alpha$) presented in Figure~\ref{fig:histogram}~(b), that non-barred,  short-barred, and A-type galaxies have relatively low $C$ (H$\ \alpha$)  values. Here the mean values of $C$ (H$\ \alpha$) for A-type and B-type are $2.2\pm0.1$ and $6.2\pm0.3$, respectively. We ran a Kolmogorov-Smirnov (KS) test on $C$ (H$\ \alpha$) to assess the probability that A-type galaxies and B-type galaxies are drawn from the same parent population. We obtain a KS probability value ($p$) of 0.006 ($<0.05$), providing evidence against the null hypothesis that the A-type and B-type $C$ (H$\ \alpha$) values are drawn from the same distribution. The Ro$C$ values of the B-type galaxies range from 0.8 to 1.3, with a mean of 1.0 (bootstrap $95\%$ CI: $0.9-1.1$), whereas those of the A-type galaxies range from 0.3 to 0.4, with a mean of 0.4 (bootstrap $95\%$ CI: $0.3-0.4$). See Figure~\ref{fig:histogram} (c) and Table~\ref{tab:002}. Finally, in Figure~\ref{fig:histogram}~(d), we colour-code the galaxies known to harbour AGNs. All A-type galaxies of small Ro$C$ actually host AGNs. The four A-type galaxies do not share a common AGN type or environment: their AGN classifications include LINER, Seyfert 1, and Seyfert 2, while three are group members and one is a cluster member (Table~\ref{tab:002}).

Given the small size of the A-type subsample ($N=4$), we additionally performed an exact Mann–Whitney U test as a rank-based cross-check. The test yields $U=0$ and a two-sided $p$-value of 0.006, confirming that the $C$ (H$\ \alpha$) values of the A-type galaxies are systematically lower than those of the B-type galaxies. We also performed an exact permutation test using the difference in the mean $C$ (H$\ \alpha$) between the A-type and B-type galaxies as the test statistic. All $\binom{11}{4} = 330$ possible assignments of four of the 11 barred galaxies to the A-type group were evaluated. The difference between the observed mean values is $\Delta \bar{C} = \bar{C}_{A} - \bar{C}_{B} = -4.0$, which lies at the extreme end of the permutation distribution, yielding a two-sided p-value of 0.006. These independent small-sample tests therefore support the result obtained from the KS test.

According to Figure~\ref{fig:C_COHA2}, Ro$C$ values close to unity in the 13 galaxies, excluding A-type galaxies, indicate that star formation and molecular gas are similarly concentrated in the centre. This is also visually evident from the similar spatial distributions of the H$\ \alpha$ and CO emission.  
In contrast, star formation in the central regions of the four A-type galaxies is relatively inactive compared to the B-type galaxies, and we find evidence that all A-type galaxies host AGNs. 
Bar-induced central gas inflow provides a plausible explanation for enhanced central gas concentration, which can further trigger co-spatial starbursts in all B-type galaxies~\citep[e.g.,][]{sakamoto1999bar,sheth2005secular}. In A-type galaxies, a similar inflow occurs, but the presence of an AGN can suppress central star formation. We elaborate on this in the following subsection.

To examine whether the low Ro$C$ values could instead be associated with the molecular gas mass fraction, we tested the correlation between Ro$C$ and  $f_\mathrm{gas}$ for the 11 A-type and B-type barred galaxies. A Spearman rank correlation test gives a correlation coefficient $ \rho=-0.04$ with $p=0.92$, indicating no evidence for a monotonic correlation between the two quantities. Thus, within the present sample, the low Ro$C$ values of the A-type galaxies do not appear to be driven simply by the global molecular gas mass fraction.

We also examined whether Ro$C$ depends on the bar length (see Table~\ref{tab:001}, $R_\mathrm{bar}/R_{25}$) for all 14 barred galaxies, including the short-barred galaxies. A Spearman rank correlation test gives $\rho=-0.12$ with $p=0.68$, indicating no evidence for a monotonic correlation between the two quantities. This suggests that the separation between the A-type and B-type galaxies is not simply driven by bar length.

Finally, we note that the analysis without masking revealed no change in the trend shown in Figure~\ref{fig:C_COHA2}, confirming that the Ro$C$ values for A-type galaxies still fall within the range of 0.3 to 0.4. This result is considered reasonable because the masking size is $\sim$ 180 pc, which is about 10 times smaller than the central region.

\subsection{The bimodal distribution of the Ro$C$ and the central suppression of star-formation by AGN feedback}

\begin{figure*}
\begin{center}
  \includegraphics[width=145mm]{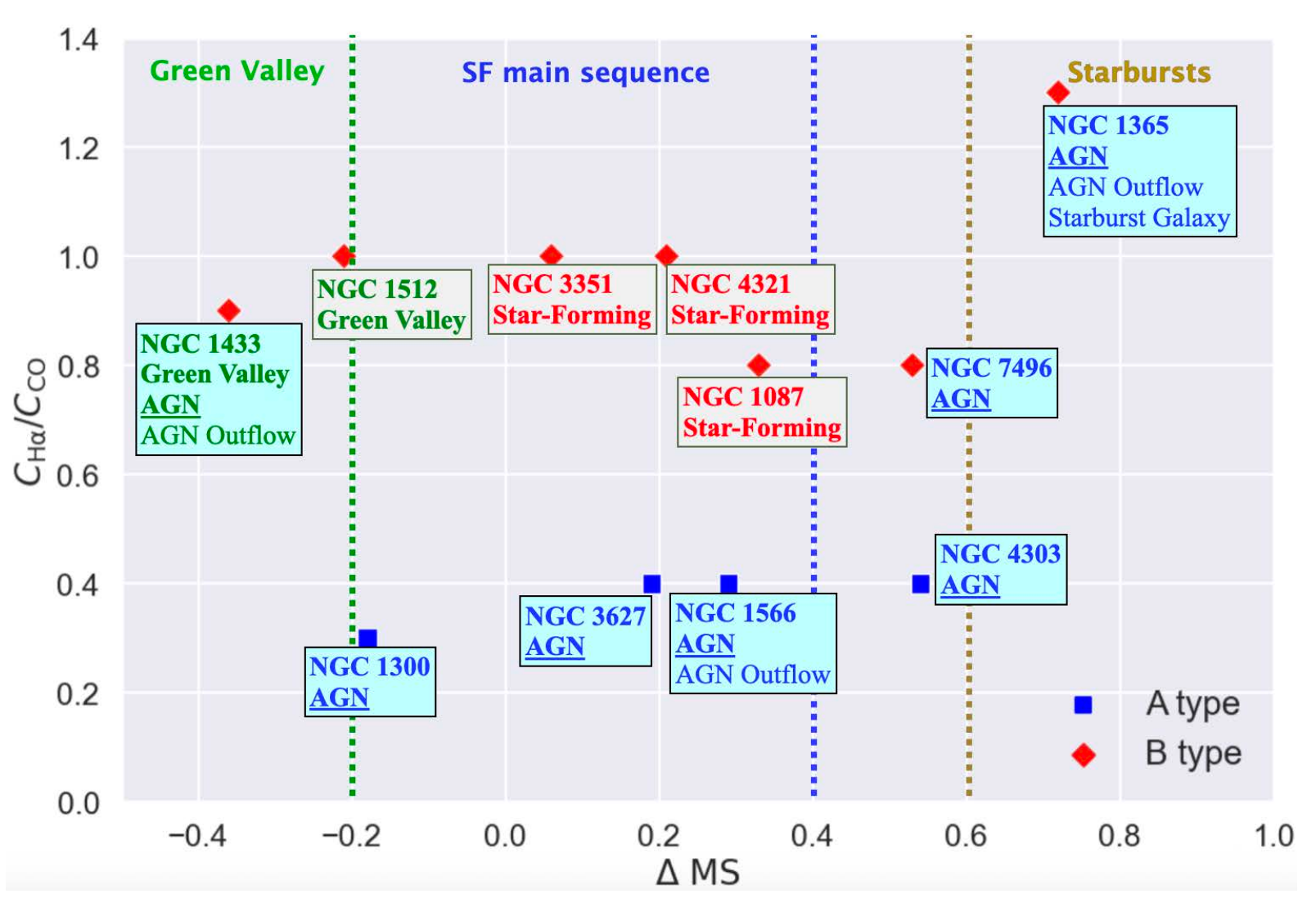}
 \end{center}
\caption{Illustrates the relationship between $\Delta$MS and Ro$C$ in 11 barred spiral galaxies. These barred spiral galaxies are distributed in bimodal, A-type, and B-type along the vertical axis (Ro$C$), and are further divided into three populations: starburst galaxies, star-forming main sequence galaxies, and green valley galaxies along the horizontal axis ($\Delta$MS). We defined the classification criteria for three galaxy populations based on and modified from~\citet{pan2024almaquest}. This panel shows that the suppression of star formation in the centres of A-type galaxies can be attributed to negative AGN feedback. Possible mechanisms include the photo-dissociation of molecular gas by intense radiation from AGN and the turbulence of molecular gas driven by AGN outflows.
}{Alt text: Illustrates the relationship between $\Delta$MS and Ro$C$ in 11 barred spiral galaxies. These barred spiral galaxies are distributed in bimodal, A-type, and B-type along the vertical axis, and are further divided into starburst galaxies, star-forming main sequence galaxies, and green valley galaxies along the horizontal axis.}
\label{fig:quenching}
\end{figure*}

In order to quantify the discussion, we define the main sequence of star-forming galaxies (MS) and also measure the offset from it ($\Delta \mathrm{MS}$). The MS is considered as one of the most valuable tools in the field of galaxy evolution, which expresses the relation between SFR and the stellar mass ($M_{*}$) in galaxies at all epochs. Equation (19) of~\citet{2019ApJS..244...24L} is used to calculate the value of $\Delta$MS. Table~\ref{tab:002} presents the results, and Figure~\ref{fig:quenching} illustrates the relationship between $\Delta$MS and Ro$C$ in 11 barred spiral galaxies. Details of the equations used are provided in Appendix A. Positive (negative) $\Delta$MS indicates a higher (lower) SFR than average. We adopt the sSFR classification of~\citet{pan2024almaquest,rodighiero2011lesser,wang2019starburst}. Three distinct populations are identified based on their offset from the main sequence: starburst galaxies, star-forming main-sequence galaxies (hereafter SFMS), and Green Valley galaxies (hereafter GV). For further details, see Appendix B.

The diversity of Ro$C$ can be attributed to the differences in SFE (${\mathrm{SFR}}/{M_{\mathrm{H_2}}}$, where $M_{\mathrm{H_2}}$ is the molecular gas mass) at the galactic centres, where A-type galaxies have lower SFE than B-types. Over half (57 per cent) of the AGN-host galaxies are classified as A-type galaxies, indicating a possible influence of AGN activity on the relative distribution of H$\ \alpha$ and CO. One possibility is that these A-type galaxies are undergoing a phase of reduced Ro$C$ values as they transition from a starburst phase to a more quiescent state. This phase may correlate with the activity level of the central black hole. Possible mechanisms include the photo-dissociation of a portion of the molecular gas by intense radiation from AGN ~\citep[e.g.,][]{saito2022kiloparsec}.

Additionally, turbulence in molecular gas clouds driven by active galactic nuclei~\citep[e.g.,][]{guillard2015exceptional} is considered a potential mechanism for the suppression of star formation. Using spatially resolved optical spectroscopic data and radio observations, \citet{nandi2025warm} suggest that AGN-driven ionised gas outflows are common across all AGN.
Otherwise, \citet{appleby2020impact} argued that from the SIMBA simulation, which uses various AGN feedback models with X-ray feedback \citep{choi2012radiative}, the AGN is important for central suppression of sSFR. A similar conclusion was reported using IllustrisTNG simulations with other models of thermal and kinetic AGN feedback~\citep[e.g.,][]{weinberger2018supermassive}.

In contrast, three B-type galaxies (NGC 1365, NGC 1433, and NGC 7496) host AGNs, indicating that the presence of an AGN alone does not necessarily result in suppressed star formation or reduced Ro$C$ values. NGC 1365 is a starburst galaxy with SFR$~\sim 20\ \mathrm{M}_{\odot}\mathrm{yr^{-1}}$. The surface density of molecular gas in the galaxy centre is approximately one order of magnitude higher than that in the other galaxies in our sample. A high concentration of the central gas ring and intense star formation suggest elevated bar-driven inflow, which may reduce the effectiveness of AGN feedback in suppressing star formation on kpc scales. However, since such inflow is expected to peak during bar formation and decline thereafter, the current state is unlikely to be long-lived~\citep[e.g.,][]{seo2013star,fanali2015bar}. In comparison, NGC 1433 is classified as a GV galaxy, with inactive star-formation in both its central and outer regions. We estimate that, as a result, the Ro$C$ value became approximately 1 (0.9). The remaining one, NGC 7496, is an SFMS galaxy, but its Ro$C$ value is moderately low (0.8), suggesting reduced star formation in its central region. For further details, see Appendix C.

Figure~\ref{fig:2model} shows schematic diagrams of an AGN host barred galaxy and a non-AGN barred galaxy, respectively. The right panel in Figure~\ref{fig:2model} shows several harsh environments for star formation. These include photo-dissociation driven by intense AGN radiation and turbulence in molecular gas clouds induced by AGN outflows. We assume that the outflow does not completely blow away the molecular gas and that high-velocity turbulence and cloud-cloud collisions may make it difficult for star formation to occur~\citep[e.g.,][]{maeda2020large}. Due to the intense radiation from AGN, ionised gas regions may dominate over a wide area in the centre. Note that this schematic diagram does not account for AGN positive feedback.

From the perspective of quenching, this result can suggest an ``inside-out quenching” pathway in AGN-barred spiral galaxies. However, AGN activity does not necessarily persist for long. It has been argued that AGN activity is short-lived on cosmic time-scales. Nevertheless,~\citet{schawinski2015active} reported that AGNs (typically lasting $10^5$ yr) can switch on and off many times during a long total growth time of $10^7-10^9$ yr estimated from~\citet{soltan1982,yu2002observational,marconi2004local}. Our research is limited to snapshots of AGN activity; the bimodality of the Ro$C$ distribution may result from AGN on-off states. Despite that, the suppression of star formation in the central region during only the AGN's active phase may have a cumulative effect through the repeated on-off cycles of AGNs, like body blows, leading to an `inside-out quenching' pathway. Furthermore, in another scenario, rapid quenching could occur during a single active AGN on-state. In either scenario, it can be suggested that the 'inside-out quenching' pathway occurs in A-type barred galaxies. The formation of bar structures causes molecular gas to collect in the central region, leading to starburst activity. However, after the starburst, the remaining gas becomes inefficient for star formation rapidly due to AGN feedback. It can mean the quenching process occurs more rapidly in AGN-barred galaxies. This point is a unique mechanism found in barred spiral galaxies and is essential to understanding galaxy evolution.

\begin{figure*}
\begin{center}
  \includegraphics[width=145mm]{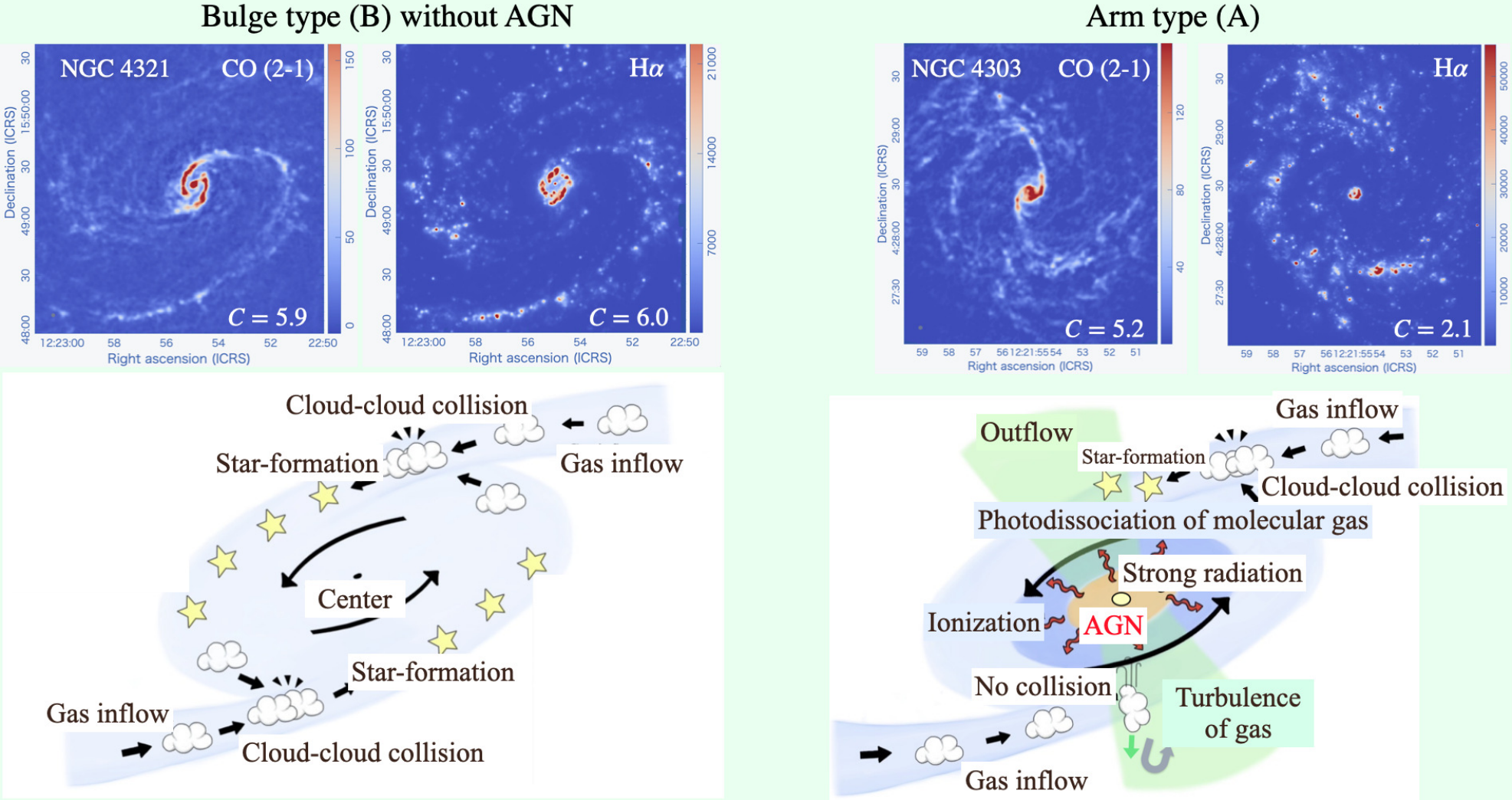}
 \end{center}
\caption{Schematic diagrams of an AGN host galaxy and a non-AGN galaxy. The right panel shows several harsh environments for star formation. We assume that the outflow does not completely blow away the molecular gas and that turbulence makes it difficult for molecular cloud-cloud collisions to occur. Due to the intense radiation from AGN, the ionised gas regions may become dominant over a wide area in the centre. Note that the right panel does not account for AGN positive feedback.}{Alt text: Schematic diagrams of an AGN host galaxy and a non-AGN galaxy. The right panel shows several harsh environments for star formation.}
\label{fig:2model}
\end{figure*}

\section{Conclusions}

We analysed 17 of 19 PHANGS survey (PHANGS-ALMA and PHANGS-MUSE) galaxies, excluding two with significant dust attenuation. We find that 13 galaxies are aligned near the one-to-one line in the $C$(H$\ \alpha$)-$C$(CO) plane. These galaxies exhibit a mean value of $0.9$ (range: $0.8-1.3$) in $C$ of H$\ \alpha$ to $C$ of CO ratio (Ro$C$). In these galaxies, there are seven B-type barred galaxies, with high $C$(H$\ \alpha$) and $C$(CO). The respective mean $C$ values are $6.2\pm0.3$ and $6.3\pm0.3$ (see Table~\ref{tab:001} and Figure~\ref{fig:C_COHA2}). 

We have found that four A-type barred galaxies exhibit star-formation suppression in their centres, with a mean Ro$C$ of $0.4$ (range: $0.3-0.4$). These four barred galaxies are all AGN host galaxies. The A-type galaxies have a high mean $C$ in molecular gas ($C=5.7\pm0.5$), but they do not have a high mean $C$ in H$\ \alpha$ ($C=2.2\pm0.1$), indicating suppressed star formation in the central region. This is consistent with a scenario in which negative feedback from AGNs suppresses star formation at the galactic centre. To facilitate a more quantitative discussion of galaxy evolution, we measured the offset from the main sequence of star-forming galaxies ($\Delta \mathrm{MS}$). In general, galaxies may evolve from the right to left (from starbursts to GV) in Figure~\ref{fig:quenching} if galaxies do not undergo gas-rich mergers. Our sample galaxies show no signs of recent mergers. The four AGN-barred galaxies are likely on a path of `inside-out quenching'. However, given the limited sample size, further research is needed to clarify the quenching pathway. In any case, our essential finding is that AGN-barred spiral galaxies exhibit a distinct specific distribution in Ro$C$ versus $\Delta$MS relationship.

We note that the activity of AGNs in these nearby galaxies is not strong enough to blow away all the molecular gas. The photo-dissociation of a part of molecular gas by intense radiation from AGN may be one of the reasons for preventing star formation. Alternatively, due to the AGN outflow, molecular gas clouds are likely to be in a high-velocity, turbulent state, which may prevent the formation of molecular cloud cores. The velocity dispersion of molecular clouds on the parsec scale remains unknown due to insufficient spatial resolution. Further investigation is needed to verify this point.

\section*{Acknowledgements}

We thank the anonymous referee for constructive and invaluable comments that improved this paper. This study makes use of the following ALMA data,
which have been processed as part of the PHANGS-ALMA CO~(2--1) survey:\\
\noindent
ADS/JAO.ALMA\#2012.1.00650.S,\par
\noindent
ADS/JAO.ALMA\#2013.1.00803.S,\par
\noindent
ADS/JAO.ALMA\#2013.1.01161.S,\par
\noindent
ADS/JAO.ALMA\#2015.1.00121.S,\par
\noindent
ADS/JAO.ALMA\#2015.1.00782.S,\par
\noindent
ADS/JAO.ALMA\#2015.1.00925.S,\par
\noindent
ADS/JAO.ALMA\#2015.1.00956.S,\par
\noindent
ADS/JAO.ALMA\#2016.1.00386.S,\par
\noindent
ADS/JAO.ALMA\#2017.1.00392.S,\par
\noindent
ADS/JAO.ALMA\#2017.1.00766.S,\par
\noindent
ADS/JAO.ALMA\#2017.1.00886.L,\par
\noindent
ADS/JAO.ALMA\#2018.1.00484.S,\par
\noindent
ADS/JAO.ALMA\#2018.1.01321.S,\par
\noindent
ADS/JAO.ALMA\#2018.1.01651.S,\par
\noindent
ADS/JAO.ALMA\#2018.A.00062.S,\par
\noindent
ADS/JAO.ALMA\#2019.2.00129.S,\par
\noindent
ALMA is a partnership of ESO (representing its member states), NSF (USA) and NINS (Japan), together with NRC (Canada), MOST and ASIAA (Taiwan), and KASI (Republic of Korea), in cooperation with the Republic of Chile. The Joint ALMA Observatory is operated by ESO, AUI/NRAO and NAOJ. The National Radio Astronomy Observatory is a facility of the National Science Foundation operated under cooperative agreement by Associated Universities, Inc. 
This study is also based on observations collected at the European Southern Observatory under ESO programs 1100.B-0651, 095.C-0473, and 094.C-0623 (PHANGS–MUSE; PI Schinnerer), and 094. B- 0321 (MAGNUM; PI Marconi), 099.B-0242, 0100.B-0116, 098. B- 0551 (MAD; PI Carollo) and 097.B-0640 (TIMER; PI Gadotti). NK was supported by JSPS KAKENHI Grant Number JP25H00661. TH was supported by JSPS KAKENHI Grant Numbers 23K22529 and 25K00020, as well as by NAOJ ALMA joint research program 2025-28A. SKS is supported by an International Research Fellowship of the Japan Society for the Promotion of Science (JSPS). TGW gratefully acknowledges support from the UK ALMA Regional Centre (ARC) Node, which is supported by the Science and Technology Facilities Council grant number ST/Y004108/1. Finally, we thank Toshiki Saito for the useful discussion regarding the use of PHANGS-MUSE data.

\section*{Data Availability}
The PHANGS-ALMA data used in this work are available at https://almascience.nrao.edu/aq/. 
The PHANGS-MUSE data are available at https://www.canfar.net/.


\bibliographystyle{mnras}
\bibliography{mnras/TY}

\vspace{10mm}

\appendix
\section{Calculation of $\Delta$MS}
 \cite{2019ApJS..244...24L} fit 15,750 galaxies using the data from $z\sim 0$ Multiwavelength Galaxy Synthesis (z0MGS), providing,

\begin{equation}
\log_{10}\mathrm{sSFR}[\mathrm{yr}^{-1]}= (-0.32)\Bigg(\log_{10}\frac{M_{*}}{10^{10}M_{\odot}}\Bigg)-10.17.
\label{eq4}
\end{equation}


This equation can be written as,

\begin{equation}
\log_{10}\mathrm{SFR}_{\mathrm{MS}}[M_{\odot}\mathrm{yr}^{-1}]= (0.68)\log_{10}{M_{*}}[M_{\odot}]-6.97.
\label{eq6}
\end{equation}
\vspace{14pt}

 The $\Delta$MS is calculated using equation (\ref{eq6}) as follows,

\begin{equation}
\Delta \mathrm{MS} \lbrack \mathrm{dex} \rbrack = \log_{10} \mathrm{SFR}[M_{\odot}\mathrm{yr}^{-1}] - \log_{10}\mathrm{SFR}_{\mathrm{MS}}(M_{*})[M_{\odot}\mathrm{yr}^{-1}],
\label{eq7}
\end{equation}

where $\log_{10} \mathrm{SFR}$ is the value of the specific galaxy obtained from observations, and $\log_{10}\mathrm{SFR}_{\mathrm{MS}}(M_{*})$ is the value on the MS line corresponding to the stellar mass $M_{*}[M_{\odot}]$ of that galaxy.

\section{Three galaxy populations along the $\Delta$MS axis}
In present-day galaxies, GV occupy an intermediate region between star-forming galaxies (the Blue Cloud) and the quiescent galaxies (the Red Sequence) ~\citep[e.g.,][]{2019ApJ...874..142K}. We define GV as systems with $9.8<\log (M_*/M_{\odot})<11.2$ and $-11.0<\log (\mathrm{sSFR/yr}^{-1}) <-10.5$, selected from the $z\sim0$ galaxies (z0MGS) of \cite{2019ApJS..244...24L}. Based on MaNGA data, \citet{belfiore2018sdss} suggest that the GV galaxies constitute a ‘quasi-static’ population undergoing a slow-quenching process. As part of the ALMaQUEST survey~\citep[ALMA-MaNGA Quenching and Star Formation Survey;][]{lin2020almaquest}, \citet{pan2024almaquest} categorized their sample into three distinct classes based on sSFR. They are; star-forming main sequence ($\log \mathrm{sSFR} > -10.5$), `moderate sSFR' ($-11.0 <\log \mathrm{sSFR} < -10.5$, GV in our definition), and `low sSFR' ($\log \mathrm{sSFR} < -11.0$). The latter two are sub-classes of quenching galaxies. These populations are represented by $\Delta$MS values of -0.2 to +0.4, -0.7 to -0.2, and -1.7 to -0.7, respectively.

We discuss below with the perspective of quenching along the $\Delta$MS axis. Present-day galaxies' star formation is thought to have progressed from the SFMS phase, through the GV phase, to the quenched phase (red sequence) over a long time-scale~\citep[e.g.,][]{walters2022quenching,pan2024almaquest}. Starburst galaxies have elevated SFRs for their stellar mass~\citep{ellison2020almaquest}, which means starbursts lie above the main sequence of star-forming galaxies, indicating high positive $\Delta$MS values. In our sample, NGC 1365 ($\log \mathrm{sSFR}=-9.76, \ \Delta \mathrm{MS}=0.72$) is a starburst galaxy whose elevated star formation may be related to central gas inflow through the bar structure. After the starburst and consumption of the available gas, galaxies are thought to undergo a quenching process~\citep[e.g.,][]{2022MNRAS.513.2850B}, observed in the decrease of the $\Delta$MS value.  

We adopt the sSFR classification of~\citet{pan2024almaquest}. However, we do not adopt their classification of `central starbursts'; instead, we define galaxies with SFR larger than four times the value in the main sequence~\citep[e.g.,][]{rodighiero2011lesser, wang2019starburst}, corresponding to $\Delta$MS $\gtrsim  0.6$ as ``starbursts”, regardless of whether starburst activity is central or not.
The ALMaQUEST data is similar to our data in the range of sSFR and stellar mass, and the star-forming main sequence line, except they lie at slightly higher redshifts ($0.02<z<0.13$). In~\citet{pan2024almaquest}, the scatter range of SFMS galaxies’ $\Delta$MS is 0.3 dex, and our sample scatter range is the same. That range is consistent with the EAGLE simulation of~\citet{matthee2019origin} at $z=0$, $M_{*}\gtrsim 3\times 10^{10}$. Thus, our sample is distinguished into three populations (`Starbursts', `Main sequence', `Green Valley') along the $\Delta$MS axis. Each of these three populations has a distinctive gas mass fraction ($f_\mathrm{gas}=M_\mathrm{H_2}/M_{*}$) range. (see Table~\ref{tab:002}). These $f_\mathrm{gas}$ values were taken from Table 3 of~\citet{yamamoto2025quantitative}. Two GV galaxies in our sample show significantly lower gas mass fractions ($f_\mathrm{gas}=1.5-1.7$ per cent), similar to those of S0 galaxies. These galaxies are still star-forming mainly in the central region, but may be quenching on a relatively short time-scale.

We note an important point from~\citet{pan2024almaquest}: as galaxies move away from SFMS to GV, the radial profile of the SFR surface density is suppressed within the central regions compared with main-sequence galaxies. That is, they have reported inside-out quenching in most ALMaQUEST galaxies, as indicated by median values. 
On the other hand, \citet{appleby2020impact} reported sSFR and gas radial profiles of SFMS and GV galaxies in the SIMBA cosmological hydrodynamic simulation. Simulations predict both inside-out and outside-in modes for galaxies with masses similar to the ALMaQUEST sample~\citep[e.g.,][]{appleby2020impact}.

\section{B-type AGN galaxies}
Three B-type galaxies (NGC 1365, NGC 7496, and NGC 1433) host AGNs, implying that the presence of an AGN does not necessarily lead to suppressed star formation or reduced Ro$C$ values, and may also increase the star formation activity. In NGC 1365, located at the top right of Figure~\ref{fig:quenching}, is a starburst galaxy with SFR$~\sim 20\ \mathrm{M}_{\odot}\mathrm{yr^{-1}}$~\citep[e.g.,][]{schinnerer2023phangs,liu2023phangs}. The surface density of molecular gas in the galaxy centre is approximately one order of magnitude greater than that of other samples. A high concentration of the central gas ring and intense star formation suggest elevated bar-driven inflow, which may reduce the effectiveness of AGN feedback in suppressing star formation on kpc scales. In addition, AGN-driven outflows have been detected in NGC 1365~\citep[e.g.,][]{venturi2018magnum}. AGN-driven outflows may locally enhance star formation by compressing dense gas, as suggested by both theoretical and observational studies~\citep[e.g.,][]{zubovas2013agn,silk2013unleashing,cresci2015magnum}. However, since such inflow is expected to peak during bar formation and decline thereafter, the current state is unlikely to be long-lived~\citep[e.g.,][]{seo2013star,fanali2015bar}. NGC 7496 is an AGN host galaxy, but it does not qualify as an A-type; the reason remains unclear, but it is important to note that its Ro$C$ is somewhat low at 0.8. Among other B-type galaxies, there are no galaxies with a Ro$C$ value of 0.8, except for NGC 1087 (NGC 1087 is a rare barred galaxy that is star-formation active in the bar, resulting in a low $C$ (H$\ \alpha$) value). See Figure~\ref{fig:C_COHA2}.

The third AGN in B-type, NGC 1433, is a GV galaxy and exhibits the most negative $\Delta$MS value in this sample. In particular, this galaxy has a low sSFR ($\log_{10} \mathrm{sSFR} = -10.82$) and extremely low star formation throughout the disc. In NGC 1433, star formation in the central region may also be suppressed by negative feedback from the AGN. Meanwhile,~\citet{combes2013alma} report that NGC 1433 exhibits AGN-driven outflows, and, as a result, AGN positive feedback may slightly increase star formation in the central region. Nevertheless, it is difficult to determine whether the central AGN exerts positive or negative feedback, as central star formation is weak. In addition, for example, compared with one of the SFMS galaxies, NGC 4321, the molecular gas mass in the outer region is only 40 per cent. Consequently, star formation actually weakens toward the outer regions, resulting in a relatively concentrated distribution of H$\ \alpha$ in the galaxy centre, which may be making Ro$C$ equal to 0.9.

\bsp	
\label{lastpage}
\end{document}